\documentclass[fleqn,usenatbib]{mnras}

\usepackage{newtxtext,newtxmath}

\usepackage[T1]{fontenc}
\usepackage{ae,aecompl}

\usepackage{graphicx}	
\usepackage{amsmath}	
\usepackage{amssymb}	

\usepackage[figure,figure*]{hypcap}

\usepackage{caption}
\usepackage{subcaption}
\title{GASP XVIII: Star formation quenching due to AGN feedback in the central region of a jellyfish galaxy}

\author[K. George et al.]{
K. George$^{1,2}$,\thanks{E-mail:koshyastro@gmail.com}
B. M. Poggianti$^3$,
C. Bellhouse$^4$,
M. Radovich$^3$,
J. Fritz$^5$, 
R. Paladino$^6$,
\newauthor  
D. Bettoni$^3$,
Y. Jaff\'e$^7$,
A. Moretti$^3$, 
M. Gullieuszik$^3$,
B. Vulcani$^3$,
G. Fasano$^3$,
C. S. Stalin$^{1}$,
\newauthor  
A. Subramaniam$^{1}$,
S.N. Tandon$^{1,8}$
\\
$^{1}$Indian Institute of Astrophysics, Koramangala II Block, Bangalore, India\\
$^{2}$Department of Physics, Christ University, Hosur Road, Bangalore 560029, India\\
$^{3}$INAF-Astronomical Observatory of Padova,  
vicolo dell'Osservatorio 5   
35122 Padova, Italy\\
$^{4}$University of Birmingham School of Physics and Astronomy,
  Edgbaston, Birmingham, England\\
$^{5}$Instituto de Radioastronomia y Astrofisica, UNAM, Campus
  Morelia, A.P. 3-72, C.P. 58089, Mexico\\ 
$^{6}$Instituto Nazionale di Astrofisica - Istituto di Radioastronomia Via P. Gobetti, 101 40129 Bologna,Italy\\
$^{7}$Instituto de F\'isicay Astronom\'ia, Universidad de Valpara\'iso, Gran Breta\~na 1111, Valpara\'iso, Chile\\
$^{8}$Inter-University Center for Astronomy and Astrophysics, Pune, India}

\date{Accepted XXX. Received YYY; in original form ZZZ}

\pubyear{2018}

\begin{document}
\label{firstpage}
\pagerange{\pageref{firstpage}--\pageref{lastpage}}
\maketitle

\begin{abstract}

We report evidence for star formation quenching in the central 8.6 kpc region of the jellyfish galaxy JO201 which hosts an active galactic nucleus, while undergoing strong ram pressure stripping. The ultraviolet imaging data of the galaxy disk reveal a region with reduced flux around the center of the galaxy and a horse shoe shaped region with enhanced flux in the outer disk. The characterization of the ionization regions based on emission line diagnostic diagrams shows that the region of reduced flux seen in the ultraviolet is within the AGN-dominated area. The CO J$_{2-1}$ map of the galaxy disk reveals a cavity in the central region. The image of the galaxy disk at redder wavelengths (9050-9250 {\AA}) reveals the presence of a stellar bar. The star formation rate map of the galaxy disk shows that the star formation suppression in the cavity occurred in the last few 10$^8$ yr. We present several lines of evidence supporting the scenario that suppression of star formation in the central region of the disk is most likely due to the feedback from the AGN. The observations reported here make JO201 a unique case of AGN feedback and environmental effects suppressing star formation in a spiral galaxy.

\end{abstract}

\begin{keywords}
galaxies: clusters: intracluster medium, galaxies: star formation
\end{keywords}



\section{Introduction} \label{sec:intro}

The strength of ongoing/recent star formation in galaxies in the local Universe is manifested in the observed distribution of galaxy colors and star formation rates. The star forming spiral galaxies populate a blue region and the S0/elliptical galaxies with little or no ongoing star formation populate the red region of the colour-magnitude distribution \citep{Strateva_2001,Baldry_2004}. Such a bimodal behaviour is also observed from the star formation rate - stellar mass relation \citep{Salim_2007}. The number density of non-star forming L$\star$ galaxies is observed to increase from $z \sim 1$ to $z \sim 0$ \citep{Bell_2004,Faber_2007}. There is a mass-dependent evolution of the number density of all and star forming galaxies between $z = 0.2$ and $4$ \citep{Ilbert_2013}. This is due to the gradual or abrupt cessation of star formation (known as quenching) in spiral galaxies. Several secular processes such as AGN/stellar feedback or the action of a stellar bar, and environmental processes such as ram pressure stripping, major mergers, harassment, starvation, strangulation are invoked to explain the star formation quenching in spiral galaxies (see \citet{Peng_2015,Man_2018}). The low mass star-forming galaxies grow in mass by star formation, keeping the number density of star-forming galaxies of a given mass quite constant by replacing quenched galaxies, as clearly required in the continuity-type analysis (see \citet{Peng_2010}).\\

In the local Universe super massive black holes with mass $>$ $10{^6}$ M\textsubscript{\(\odot\)} at the center of galaxies are closely linked to galaxy formation and can also influence their evolution  \citep{Silk_1998,Magorrian_1998,Gebhardt_2000,Ferrarese_2000} (see \citet{Kormendy_2013} for a review). This is possible since the black hole at the center accrete copious amounts of gas present in the disk making the galaxy go through an active galactic nucleus (AGN) phase \citep{Bondi_1952,Gaspari_2013,Tremblay_2016}. The exact nature of the mechanism for gas accretion and the AGN phase are not fully understood, and various mechanisms could be effective, including major mergers \citep{Sanders_1988a}, internal instabilities \citep{Hopkins_2009} and enhanced activity due to ram-pressure by the intra-cluster medium \citep{Poggianti_2017a}. The kinetic and radiative energy from the accreting black hole (in the form of radiative heating, outflow and jet) can ionize the cold gas near its vicinity, thus changing the dynamical state of the gas influencing the conditions necessary for star formation \citep{Hopkins_2006,Heckman_2014}. In extreme cases, the gas can be expelled from the galaxy halting further star formation. This is hypothesized to be one of the channels for converting a star forming galaxy into a non star forming, quiescent galaxy \citep{Dimatteo_2005,Cheung_2016}. The same molecular hydrogen gas responsible for star formation can also be accreted by the black hole, hence star formation and AGN activity are usually tightly coupled at galaxy centers. The AGN thus can have a negative impact with the dual role of suppressing both star formation and gas accreting onto the black hole and this process is refereed to as AGN feedback (see \citet{Fabian_2012} for a recent review).\\

Star formation in spiral galaxies can be suppressed also by  stellar bars \citep{Masters_2010,Masters_2012,Cheung_2013,Gavazzi_2015,Hakobyan_2016,James_2016,Spinoso_2017,Khoperskov_2018,James_2018}. The presence of a stellar bar in massive star forming galaxies has been argued to be a dominant process in  mass dependent star formation quenching  and in regulating the redshift evolution of specific star formation rates for field galaxies \citep{Gavazzi_2015}. The likelihood for disk galaxies hosting stellar bar is found to be anti-correlated with specific star formation rate regardless of stellar mass and the prominence of the bulge \citep{Cheung_2013}. The presence of stellar bars can quench star formation in the central regions of galaxy by suppressing the star formation along the co-rotation radius of the bar \citep{James_2018}. The shock and shear generated within the galaxy due to the presence of a bar can create turbulence preventing the molecular gas from collapse thereby inhibiting star formation \citep{Reynaud_1998}. The stellar bar in a galaxy can also dynamically re-distribute the gas making the region close to the bar devoid of fuel for further star formation \citep{Combes_1985}.\\

Star forming galaxies in the dense environments of galaxy clusters are subject to other forms of star formation quenching such as ram-pressure stripping, strangulation and harassment \citep{Boselli_2006}. Ram-pressure stripping by the intra-cluster medium is an efficient way of removing gas from infalling galaxies \citep{Gunn_1972}.  In some cases, stars can form in the stripped gas giving the appearance of a jellyfish at optical or UV wavelengths \citep{Cortese_2007,Smith_2010,Owers_2012,Ebeling_2014,Fumagalli_2014, Poggianti_2017b,Poggianti_2019}. There is observational evidence for AGN and ram-pressure stripping operating separately, quenching star formation in galaxies \citep{Wylezalek_2016,Boselli_2006}, and recently a possible connection between these two phenomena has been established \citep{Poggianti_2017a}.\\

We report unprecedented observations of a galaxy undergoing intense ram-pressure stripping and at the same time experiencing star formation quenching in the central region. Our multi-wavelength dataset supports the notion that the suppression of star formation is due to the presence of an accreting black hole via feedback processes in the central 8.6 kpc.

\subsection{The jellyfish galaxy JO201}

The galaxy JO201 is one of the most extreme cases of ram-pressure stripping in action and has been studied in detail for H$\alpha$ kinematics, presence of an AGN, molecular gas content and ongoing star formation \citep{Bellhouse_2017, Poggianti_2017a,Moretti_2018a,George_2018} as part of the GASP survey \citep{Poggianti_2017b}. GASP (GAs Stripping Phenomena in galaxies with MUSE) aims at investigating the gas removal process in a sample of 114 disk galaxies at redshifts 0.04-0.07, using the spatially resolved integral field unit spectrograph MUSE \citep{Poggianti_2017b}.
This program focuses on galaxies in various stages of ram pressure stripping in clusters \citep{Jaffe_2018,Vulcani_2018c}, from pre-stripping (undisturbed galaxies of a control sample), to initial stripping, peak stripping \citep{Bellhouse_2017,Gullieuszik_2017, Poggianti_2017b,Moretti_2018b}, and post- stripping \citep{Fritz_2017}, and passive, and on a number of physical processes in groups and filaments ranging from stripping to gas accretion, mergers, and cosmic web \citep{Vulcani_2017,Vulcani_2018a,Vulcani_2018b,Vulcani_2018c}.\\

JO201\footnote{$\alpha$(J2000) = 00:41:30.325, $\delta$(J2000) = - 09:15:45.96} with a spectroscopic redshift $z \sim$ 0.056 is located at a luminosity distance of $\sim$ 250  Mpc in the Abell 85 galaxy cluster \citep{Moretti_2017}\footnote{The angular scale of 1" corresponds to 1.087 kpc at the Abell 85 galaxy cluster rest frame.}. The galaxy is of spiral morphology with a total stellar mass $\sim$ 3.55 $\times$ $10^{10}$ M$_{\odot}$ \citep{Bellhouse_2017}. The galaxy JO201 is falling into Abell 85 from the back along the line of sight with a slight inclination to the west, hosting intense star formation in the disk and in the stripped material due to the effect of ram-pressure stripping compressing the gas \citep{Bellhouse_2017,George_2018}. The galaxy's high velocity within the cluster (3363.7 km/s with respect to the mean velocity of Abell 85) and its proximity to the cluster centre make it an extreme case of ram pressure stripping. The presence of an AGN in JO201 and in other five out of seven jellyfish galaxies with long tails of stripped gas supports the idea that the AGN is triggered by intense ram-pressure stripping, which can potentially funnel gas into the central parts of the galaxy \citep{Poggianti_2017a}.\\

The stellar populations in the galaxy disk of JO201 consist of both younger and older populations, the relative contributions of which are difficult to disentangle from optical observations. The UV flux is coming from the stellar photospheres of young stars and directly traces the star formation over the past 100-200 Myr \citep{Kennicutt_2012}, while the optical flux at redder wavelengths traces more evolved stellar populations. The H$\alpha$ emission on the other hand is due to star formation on timescales of 
10-20 Myr at most. The ongoing star formation in the disk of the jellyfish galaxy JO201 has been studied using UV and H$\alpha$ data in \citet{George_2018}. This paper builds on the results presented in \citet{Poggianti_2017a,George_2018,Bellhouse_2019} and focuses on the star formation properties in the central region surrounding the AGN in JO201, combining MUSE emission lines, optical red continuum (9050-9250 {\AA}), UVIT UV data and ALMA CO map for the J$_{2-1}$ transition.\\

We discuss the observations in section 2, and present the results in section 3, discussion in section 4. We summarize the key findings from the study in section 5. Throughout this paper we adopt a Salpeter 0.1-100 $M_{\odot}$ initial mass function, and a concordance $\Lambda$ CDM cosmology with $H_{0} = 70\,\mathrm{km\,s^{-1}\,Mpc^{-1}}$, $\Omega_{\rm{M}} = 0.3$, $\Omega_{\Lambda} = 0.7$.


\begin{figure}
\includegraphics[width=8.0cm]{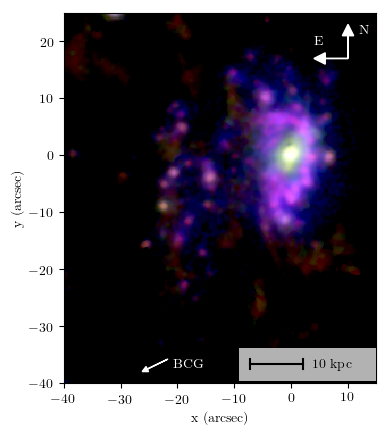}
\caption{Color composite image of J0201 made from combining NUV (colored blue), $\mathrm{H}{\alpha}$ (colored red) and [OIII] (colored green). The direction of the brightest cluster galaxy (BCG) is shown by the arrow. Note the main disk of the galaxy with intense star formation (as seen in NUV and $\mathrm{H}{\alpha}$) and the knots of star formation in the stripped material from the galaxy. The center of the galaxy disk and the region around are dominated by [OIII] emission due to the accreting black hole at the center.}\label{figure:JO201rgb}
\end{figure}

\section{Data and Observations} \label{sec:data}

The galaxy JO201 was observed at optical wavelengths as part of the WINGS and OmegaWINGS surveys \citep{Fasano_2006,Gullieuszik_2015,Moretti_2017} and with the MUSE integral-field spectrograph on the VLT under the programme GASP with photometric conditions and image quality of $\sim$ 0\farcs7 FWHM, as described in detail in \citet{Bellhouse_2017}. The MUSE observations cover the main body and the stripped tails of the galaxy. The galaxy is classified as a Seyfert 2 AGN based on the emission line spectra \citep{Poggianti_2017a}. The analysis of MUSE data reveals extended ionized gaseous emission out to $\sim$ 60kpc from the stellar disk of the galaxy with kinematics indicative of significant stripping  in the line-of-sight direction \citep{Bellhouse_2017}. Star formation was detected from both $\rm H\alpha$ emission and UV imaging in the disk and in the tail \citep{Bellhouse_2017,George_2018,Bellhouse_2019}.

The emission line fluxes from the spectrum of each MUSE spaxel are first corrected for stellar absorption using the best fitting combination of single stellar population models to the MUSE spectra using the SINOPSIS code \citep{Fritz_2017}. The emission lines are then fitted with models comprising single or double Gaussian profiles using kubeviz \citep{Fossati_2016} (see \citet{Bellhouse_2017} for details). The galaxy has a larger line-of-sight component causing the emission lines in certain regions to be non-gaussian in nature which required a double component fit. The double component fits are used in any given spaxel if the two components were detected to S/N $>$ 3 and separated in velocity, else the single-component fits with S/N $>$ 3 were used.  The primary component is either the single component fit or the narrower component  of the double component fit. The emission line fluxes measured from the primary component are used to create the flux maps of the galaxy. 

JO201 was observed in FUV (F148W filter, $\lambda_{mean}$=1481 {\AA}, $\delta\lambda$=500 {\AA}) and NUV (N242W filter, $\lambda_{mean}$=2418 {\AA}, $\delta\lambda$=785 {\AA}) wavelengths using the UVIT instrument on board the Indian multi-wavelength astronomy satellite ASTROSAT \citep{Agrawal_2006,Tandon_2017}. The UVIT imaging yields a resolution of $\sim$ 1.2" for the NUV  and $\sim$ 1.4" for the FUV channels. The details of the UV observations of JO201 are given in Table 1 of \citet{George_2018}.

The UV imaging data along with the  $\mathrm{H}{\beta}$ (4861.33 {\AA}), [OIII] (4958.91 {\AA}, 5006.84 {\AA}), [FeVII] (6086.97 {\AA}), [NII] (6548.05 {\AA}, 6583.45 {\AA}), $\mathrm{H}{\alpha}$ (6562.82 {\AA}) and [SII] (6716.44 {\AA}, 6730.81 {\AA}) emission line flux maps of JO201 are used in this study. We note that the NUV and FUV images of the JO201 galaxy disk show very similar features. The NUV image has a better spatial resolution than the FUV hence we use the NUV image to probe ongoing and recent star formation in the galaxy disk.   

JO201 was observed with ALMA in Cycle 5, using Band 3 (100 GHz) and Band 6 (230 GHz) to observe the CO (J$_{1-0}$)  and CO (J$_{2-1}$) transitions, respectively. A full description of these observations and results is given in Moretti et al. (in preparation), here we only summarize the most salient aspects of the data. Mosaics to cover the full disk and the tails have been obtained. The ALMA configurations used provide a resolution of $\sim$ 1" in both bands, and allow to recover spatial scales up to 20 and 10 " in band 3 and 6, respectively. The data have been calibrated using the standard procedure (Pipeline-CASA51-P2-B) and imaged with the task clean using CASA version 5.4. The RMS achieved in 20 km/s wide channels have been
0.5 and 0.85 mJy/beam, for CO(1-0) and CO(2-1) respectively, using weighting Briggs with robust 0.5.\\

\section{Results}

\subsection{Star formation cavity in the JO201 galaxy disk}

\begin{figure*}
\centerline{\includegraphics[width=12cm]{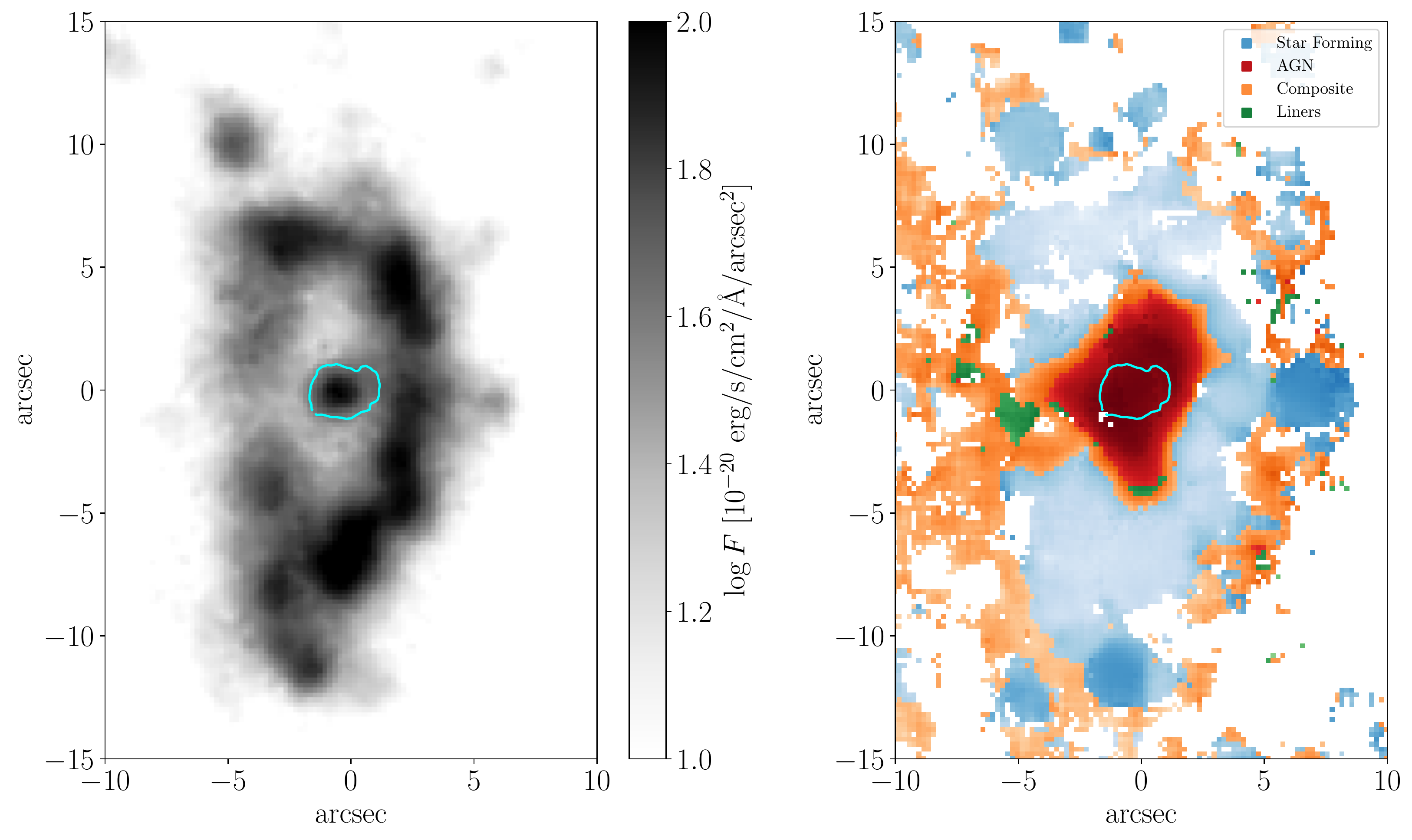}}
\caption{JO201 galaxy disk NUV image (left panel) along with regions dominated by AGN, composite (AGN+SF) and star formation (right panel). 
The NUV image is showing enhanced emission at the center and along a horse shoe shaped region in the disk of the galaxy. 
The AGN dominated region is marked in red color. The composite (AGN+SF) region, marked in orange color, occupies a thin rim around the AGN dominated region and inside the star forming region marked in blue color. The region occupied by LINER emission is marked in green. The [FeVII] 6086.97 {\AA} emission line region at the center is shown by the cyan color contour.}\label{figure:JO201} 
\end{figure*}

\begin{figure}
\centerline{\includegraphics[width=8cm]{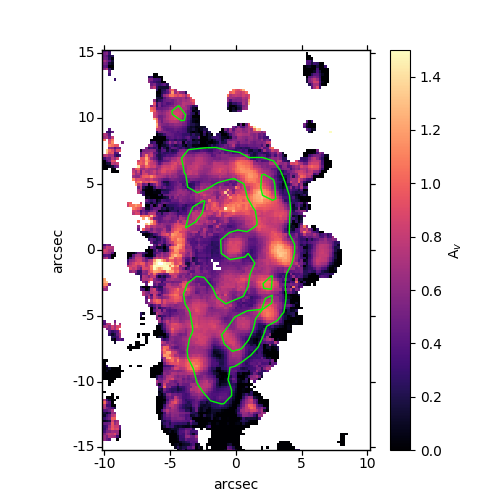}}
\caption{The V-band extinction map of the galaxy disk (A$_{v}$ in magnitude) derived from the Balmer decrement (H$\alpha$/H$\beta$=3.1) based on MUSE observations. The region near the center where we are seeing reduced UV flux is not  having a high extinction compared to the outer disk of the galaxy with enhanced UV flux. The green contour is taken from the NUV image shown in Fig \ref{figure:JO201}.
}\label{figure:JO201Av}
\end{figure}
\begin{figure*}
\centerline{\includegraphics[width=14cm]{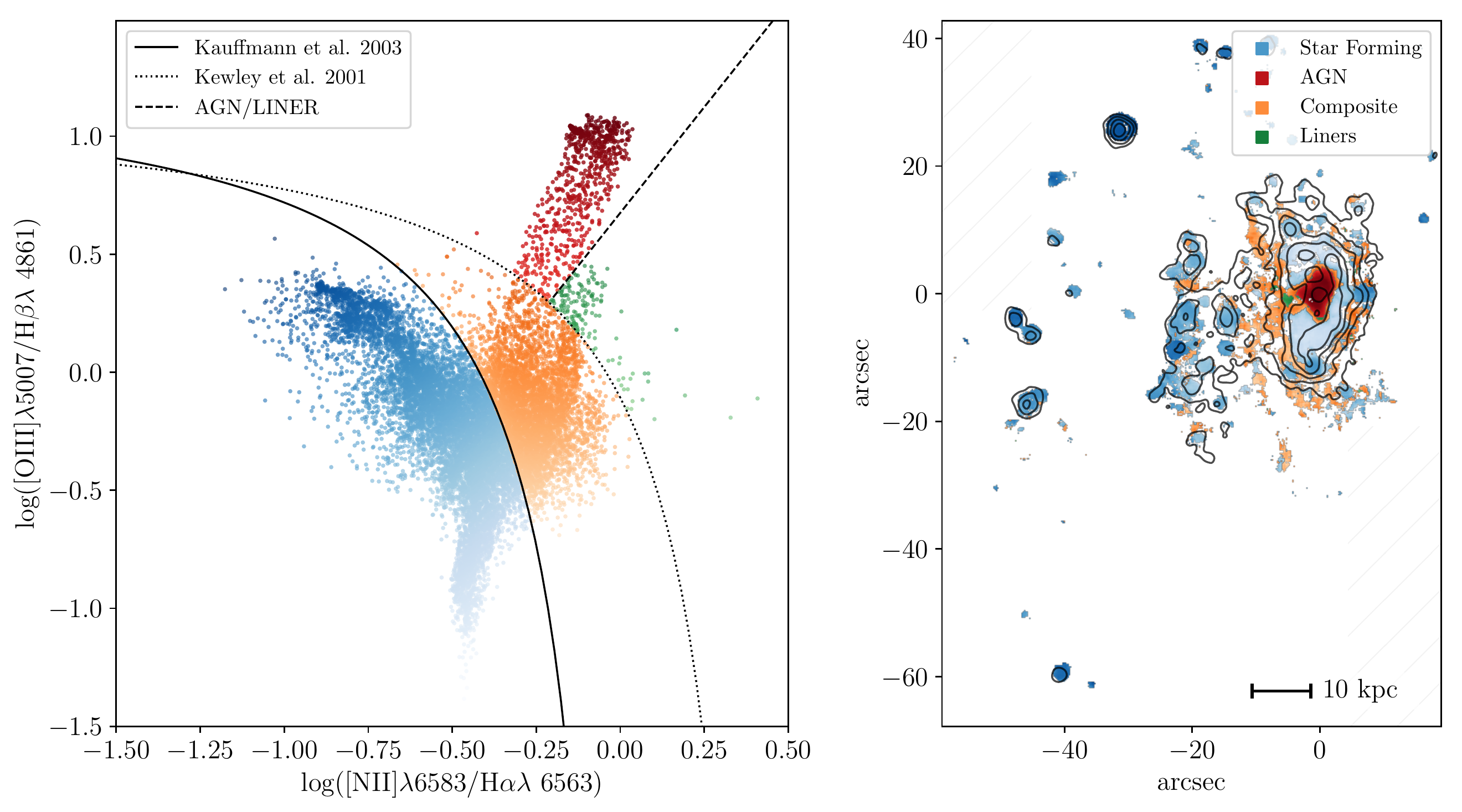}}
\caption{The BPT line-ratio diagnostics for JO201, mapped on top of the galaxy (right). The first (dominant) component of the double component fit from \citep{Bellhouse_2019} is plotted. The black lines separating different ionisation sources on the left panel come from \citet{Kauffmann_2003}, to separate star formation dominated regions from composite; \citet{Kewley_2001}, to separate composite from AGN/LINER regions; and the AGN/LINER separator, taken from \citet{Sharp_2010}. The contours on the right panel correspond to the stellar FUV image contours. The distribution of the ionized gas in the galaxy disk shows an AGN region in centre of the galaxy surrounded by  regions of ongoing star formation.}\label{figure:JO201bpt}
\end{figure*}

Fig \ref{figure:JO201rgb} shows the color-composite (RGB) image of JO201 made from combining the NUV image (blue), $\mathrm{H}{\alpha}$ (red) and [OIII] (green) emission line maps.  The galaxy disk and the stripped material show NUV and $\mathrm{H}{\alpha}$ emission due to the presence of ongoing star formation (see also Fig \ref{figure:JO201bpt} .  The western region of the disk is the first contact point of the galaxy with the hot intra-cluster medium of the Abell 85 galaxy cluster \citep{Bellhouse_2017,George_2018}. There the ram-pressure compresses the gas and induces enhanced star formation, as demonstrated by the UV and $\rm H\alpha$ enhancement along a horse shoe shaped region to the west of the center. Moreover, there is a region surrounding the center of the disk that is dominated by [OIII] emission which interestingly has a reduced NUV and $\mathrm{H}{\alpha}$ flux. The left most panel of Fig \ref{figure:JO201} shows the reduced UV flux region 
surrounding a central region with UV emission
(we postpone the discussion on the other panel  to later in this section). The contours created from this NUV image are used in the rest of our analysis to identify star forming regions as well as the region with reduced star formation on the disk of JO201.
We investigated whether the reduction in UV flux could be due to dust extinction (the map is shown in Fig \ref{figure:JO201Av}), and concluded this is not the case. The A$_{v}$ map is created from the MUSE spectra using the Balmer decrement, assuming an intrinsic ratio H$\alpha$/H$\beta$ = 3.1 typical of regions ionized purely by AGN \citep{Osterbrock_2006}. In fact, the A$_{v}$ values are generally low throughout the central region while extinction is higher along the horse shoe shaped region seen in NUV.

Emission line diagnostic diagrams \citep{Baldwin_1981} can be used to get clues on the mechanisms of gas ionization as a function of the position within the galaxy. The $\mathrm{H}{\alpha}$, [SII], [OIII] and [NII] emission line flux maps for JO201 obtained from the GASP MUSE data are used to create the line diagnostic diagrams, based on which the contribution from star formation, composite (AGN+SF) and AGN are identified as shown in Fig \ref{figure:JO201bpt} (\citet{Bellhouse_2019}, \citet{Poggianti_2017a}). We show in Fig \ref{figure:JO201} the regions corresponding to AGN (red), Composite (AGN+star formation, orange) and star formation (blue) overlaid over the NUV image of the disk of JO201. 

The presence of hot gas in the central region can be studied from ionization lines of Fe. The [FeVII] 6086.97 \AA $\,,$ emission at the center of the galaxy is shown with cyan color contour of level 10 \% of the peak value at the center in Fig \ref{figure:JO201} (the emission line profile is displayed \citet{Radovich_2019}). The [FeVII] emission region corresponds to the central bright NUV source. We note that the AGN contributes to and possibly dominates the UV flux at the galaxy center. Reduced star formation can be present at that location but this may not be revealed as the diagnostic diagrams are dominated by the AGN.
The [FeVII] line can be due to the energy output from the AGN at the center of the galaxy. The detection of this and higher ionization lines (e.g. [FeX] 6374 {\AA}) is generally explained \citep{Mazzalay_2010} with the presence of hot gas ($T > 10^5$K)  heated either by the  AGN continuum, or e.g. related to shocks triggered by radio jets \citep{Axon_1998}. This hot gas component may contribute to the observed central UV emission via free-free or free-bound radiation processes \citep[see e.g.][]{Munoz_2009}.  Furthermore, there is also a thin UV connection between the central source and the galaxy disk, and in correspondence to this there is a significant decrease in UV flux. 

The most striking result from Fig \ref{figure:JO201} is that the NUV image clearly shows a region around the center that has reduced flux compared to the horse shoe shaped region on the western side of the disk that hosts instead intense star formation. The FUV image also shows a similar morphology as shown in Fig.~4 of \citet{George_2018}. Further outward, there is intense UV emission coming from the northern, western and southern regions of the disk where especially on the western side there appears to be enhanced star formation. As shown in Fig \ref{figure:JO201} (also see Fig \ref{figure:JO201bpt}), the classification based on line diagnostic diagram demonstrates that the central source (diameter $\sim$ 2.71" $\sim$ 3kpc) and its close surroundings (diameter $\sim$ 7.88" $\sim$ 8.6 kpc) are dominated by the emission due to the AGN. The composite region occupies a rather thin rim between the AGN region and the star formation dominated region. This is further extended to the east ward of the galaxy, with no corresponding UV emission, as clear from the NUV image. This could be explained by the existence of shocks dominating the region covered by composite emission, that also extend around the horse shoe shaped region. The AGN+Composite region coincides with the region of reduced UV flux.


The region classified as star forming is instead very well matched with the horse shoe region seen in the NUV image. Thus, Fig \ref{figure:JO201} shows that the AGN and composite regions are physically separated from the enhanced star forming region on the galaxy disk seen both in NUV and $\rm H\alpha$. 

Thus, the UV imaging clearly shows a cavity surrounding the central AGN source and an outer disk region with enhanced star formation. We found that there is a factor of 2 change in surface brightness between the outer disk and the cavity (excluding the central source that could be contaminated by AGN) of JO201. This can in effect be translated to the relative change in the star formation rate density between the disk and the cavity of JO201. The star formation rate is computed for a Salpeter initial mass function from the FUV luminosity (L$_{FUV}$) \citep{Kennicutt_1998} and using the form of equation as described in \citet{Iglesias_2006}, adopted in \citet{Cortese_2008} and shown in \citet{George_2018}.  Note that the formula is derived using $Starburst99$ synthesis model \citep{Leitherer_1999} for solar metallicity and a Salpeter 0.1-100 $M_{\odot}$ initial mass function.
The extinction correction is performed to the FUV luminosity using the method described in section 3.4 of \citet{George_2018}. The integrated star formation rate density (SFR/Area) of the disk region (as defined by the UV contour shown in green in figures) is found to be 0.39 M$_{\odot}$/yr/kpc$^2$ and for the cavity (as seen in Fig \ref{figure:JO201}) to be 0.14 M$_{\odot}$/yr/kpc$^2$. There is a factor $\sim$ 2.7 drop in star formation between the cavity and the disk region of JO201.





We emphasize here that while the MUSE data demonstrate that the ionized gas emission in the central region of the disk is dominated by AGN ionization we cannot exclude the presence of even a significant level of star formation, but the UVIT observations confirm that star formation in that region is significantly suppressed with respect to the horse-shoe shaped region.\\

The existence of the UV cavity, the presence of the AGN and the coincidence between the AGN-dominated ionized region and the UV cavity strongly suggest that the cause for the suppression of the star formation in the central region is feedback from the AGN. The central AGN can release the energy in the form of outflow, jet and heat which ionize the gas in the disk of the galaxy. We checked for signatures of a possible outflow related to the AGN in the MUSE velocity map of the central region of the disk. We detect narrow and broad emission line components in the central AGN region \citep{Bellhouse_2017,Poggianti_2017a}: as discussed in more detail in a separate paper (\citet{Radovich_2019}), the broad component appears to be related to an outflow.

\begin{figure}
\centerline{\includegraphics[width=8cm]{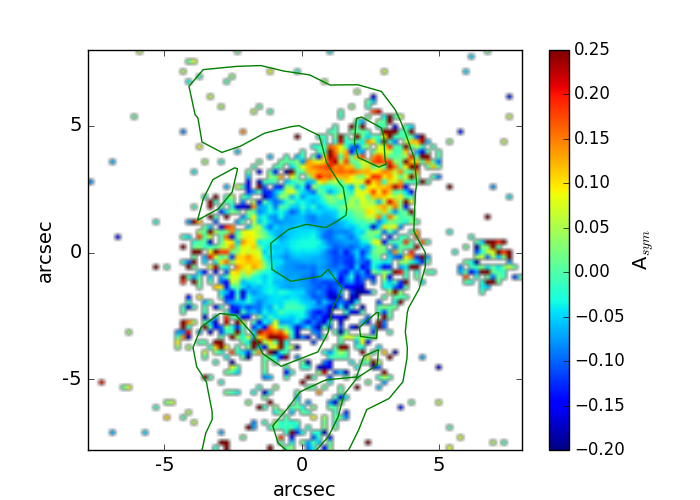}}
\caption{The asymmetry map of the [OIII] emission line on the disk of JO201 is shown with the NUV contours (green colour) overlaid. The regions with larger line asymmetry are tracing the boundary of the UV cavity, and could be tracing a spherical outflow or a bubble propagating into the medium from the AGN which could be possibly suppressing the star formation in the cavity.}\label{figure:JO201bubble}
\end{figure}

Evidence for a possible impact of the AGN on a larger scale comes from 
the asymmetry map of the [OIII] 5007 {\AA} emission line which is shown together with the NUV contours in Fig \ref{figure:JO201bubble}. The line asymmetry, $A_{\rm sym}$ \citep{Whittle_1985}, is based on the velocities measured at 10\%, 50\% and 90\% of the cumulative flux percentiles ($v_{\rm 10}$, $v_{\rm 50}$ and $v_{\rm 90}$). We adopted the definition in  \citet{Liu_2013} (see also \citet{Radovich_2019} for details): $A_{\rm sym} = \frac{(v_{\rm 90} - v_{\rm 50}) - (v_{\rm 50} - v_{\rm 10}) }{v_{\rm 90} - v_{\rm 10}}$.  In this definition,  positive/negative values of $A_{\rm sym}$ indicate red/blue  asymmetric lines.
Note that the regions on the galaxy disk with larger line asymmetry (redder/yellow regions) are mostly tracing the boundaries of the UV cavity, and could be tracing a larger spherical outflow or a bubble propagating into the medium from the AGN. This could possibly be the cause for the suppression of star formation in the UV cavity. \\

\begin{figure}
\centerline{\includegraphics[width=8cm]{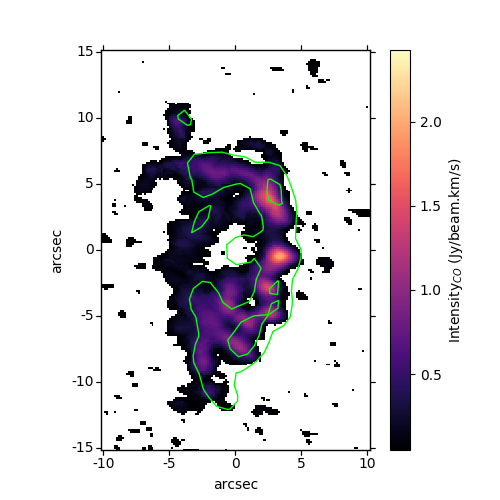}}
\caption{The CO J$_{2-1}$ transition intensity map of JO201 in flux units of Jy/beam.Km/s. The green contour is taken from the NUV image shown in Fig \ref{figure:JO201}. Note the cavity in CO map around the central AGN.}\label{figure:JO201alma} 
\end{figure}

The CO J$_{2-1}$ transition intensity map made from the ALMA observations of JO201 disk is shown in Fig \ref{figure:JO201alma}. The CO map (which traces the cold phase of molecular gas) is clearly showing a region with no detection around the central region of the galaxy. 
This observational result can be interpreted in the context of AGN feedback; the energy from the AGN is sweeping out or ionising the medium around the central region of the galaxy leaving no molecular hydrogen but instead ionized hydrogen as revealed from H$\alpha$ imaging observations. The absence of molecular hydrogen leads to a halt in ongoing star formation since the start of the AGN activity \citep{Cicone_2014}. The coincidence of the location of the cavity seen in the UV and CO data (with CO cavity sitting inside the UV cavity) confirms our hypothesis that the energetic feedback from the AGN should have suppressed the star formation in the central region of JO201 disk. 


\subsection{The stellar bar in JO201}

\begin{figure}\centering
\centerline{\includegraphics[width=8.0cm]{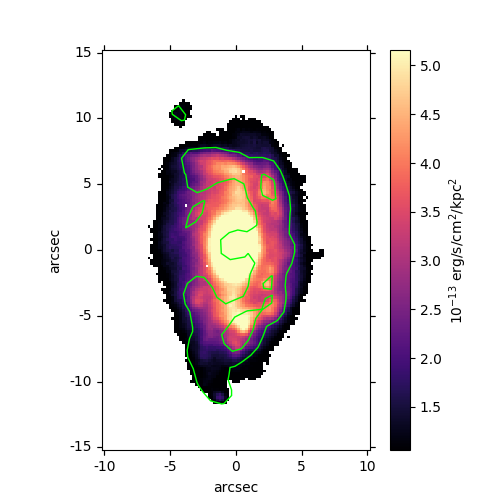}}
\caption{The MUSE image of the disk of JO201 created integrating 9050-9250 {\AA} wavelength slice. The stellar bar like feature is clearly seen in the disk of galaxy. NUV contours are overlaid (in green) to highlight the cavity with reduced UV flux.}\label{figure:JO201bar} \end{figure}

\begin{figure}\centering
\centerline{\includegraphics[width=8.0cm]{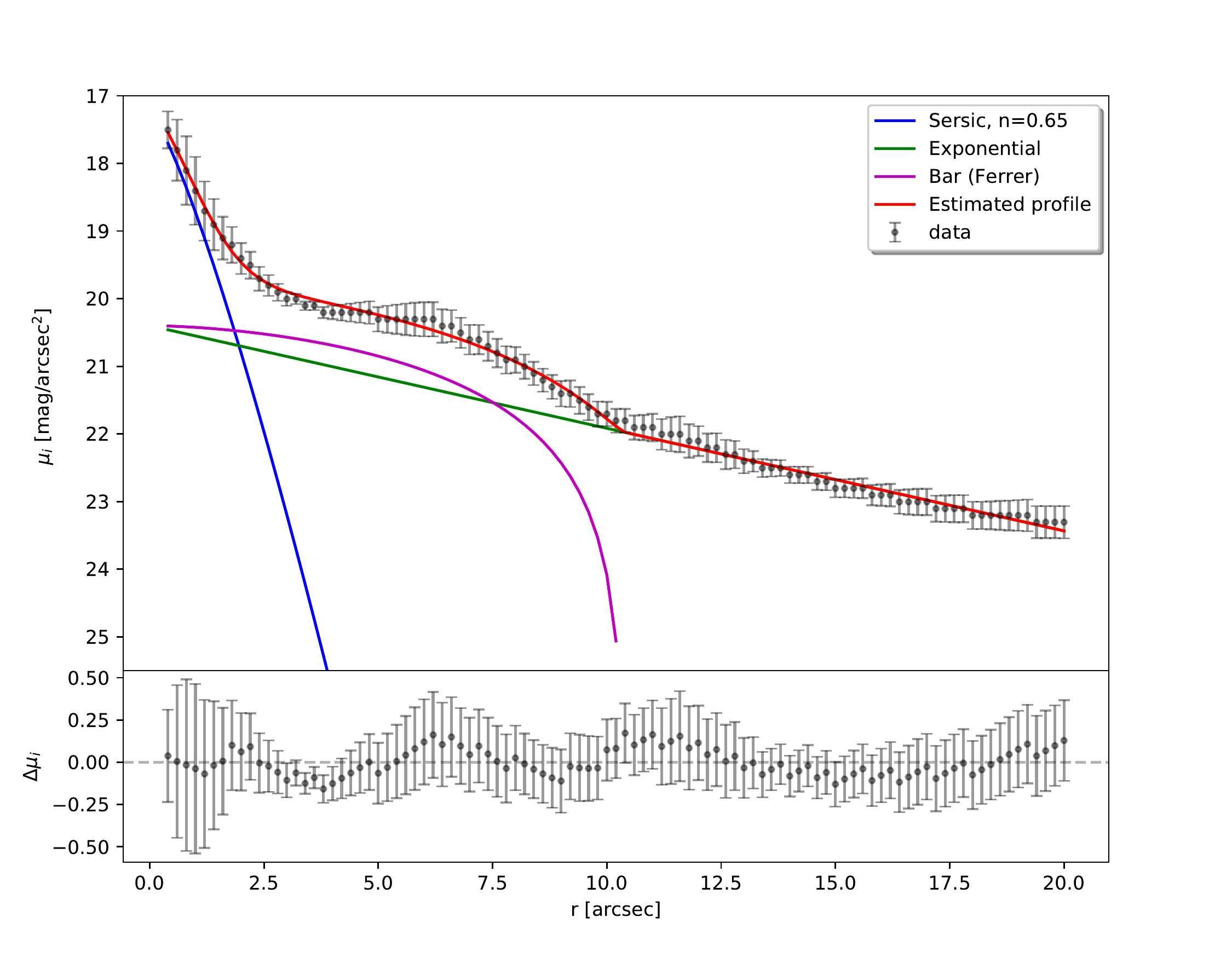}}
\caption{The stellar profile fit to the MUSE $i$ band image of the disk of JO201. The Sersic (blue), Exponential (green) and Ferrer (cyan) function fit to the light distribution is shown. Sersic function is for the central bulge, exponential for the disk and the Ferrer function is used to fit the stellar bar in JO201.}\label{figure:JO201barfit} \end{figure}

The JO201 image observed from 9050-9250 {\AA} MUSE data reveals a stellar bar like feature oriented in the North-South direction (Fig \ref{figure:JO201bar}). The stellar bar is of length $\sim$ 13 kpc. Stellar bars are known to be sometimes able to suppress star formation in the disk of galaxies \citep{Masters_2010,Masters_2012,Cheung_2013,Gavazzi_2015,James_2016,Spinoso_2017,Khoperskov_2018,James_2018}. The kinematic signature of the bar is usually probed using the Calcium II triplet lines at 8498, 8542, and 8662 {\AA}. The Ca II triplet absorption lines trace evolved stellar populations (particularly due to the contribution from stars in red giant branch (RGB) phase) \citep{Jones_1984,Armandroff_1988}.  The MUSE data are strongly affected by sky emission at red wavelengths, and therefore in the usual analysis of GASP galaxies the kinematic analysis is performed only up to the H$\alpha$ region. However, in the present case we have decided to analyze the redder region (from 8350 {\AA} restframe), after having subtracted the sky emission using the ZAP code \citep{Soto_2016}. In order to do this, we have also used the E-MILES stellar libraries extended in the red region of the spectra \citep{Vazdekis_2010,Vazdekis_2016}. The stellar kinematic map of JO201 however is not showing the presence of velocity structures expected from the presence of a stellar bar (see also Fig 10 of \citet{Bellhouse_2017}. This probably means that even though visible at red wavelength, the stellar bar is very faint and is not able to alter significantly the kinematics. For this work we also performed a two dimensional multi-component fit to the $i$ band image of the galaxy. To derive the luminosity profile we used the ellipse task in the isophote IRAF package \citep{Jedrzejewski_1987}. The resulting profile was then fitted with a three component model: a Sersic \citep{Sersic_1963}, an exponential disk \citep{Freeman_1970} and a modified Ferrer law for the bar \citep{Peng_2010}. 
Fig \ref{figure:JO201barfit} presents the multi-component (Sersic,exponential,Ferrer) fit to the light profile and clearly shows the presence of the stellar bar. The outer truncation radius of the fitted Ferrer function is  10.65" (11.57 kpc). 

Importantly, the stellar bar is visible only in the redder wavelength (above 7000 {\AA}) optical image of JO201 disk which should be tracing the flux from old stellar populations. The fact that the bar is only composed of old stars and does not host any recent star formation is also proven by the lack of UV emission tracing the bar. The stellar bar in JO201 is long and comparable to the bar length of similar mass disk galaxies in the local Universe \citep{Hoyle_2011}. The length of the bar further supports the notion that the bar is several Gyr old, as longer bars require a long timescale to form \citep{Gadotti_2006}. Since the galaxy is on first infall into the cluster, close to pericenter (see discussion in \citet{Bellhouse_2017}), and given the cluster crossing time, the old stellar bar appears to be a remnant of the secular evolution prior infall into the cluster, rather than tidally induced within the cluster environment \citep{Lokas_2016}. We also note that the CO map also does not show any evidence of bar, confirming there is no young bar present in the galaxy.

\subsection{Star formation history maps of the JO201 disk}

Important clues about the origin of SF suppression in the cavity can be obtained from stellar ages considerations in the various regions. The spectrophotometric fitting code SINOPSIS is used to derive the star formation rate (SFR) map of JO201 disk at different lookback times from the MUSE spectral data (SINOPSIS: \citet{Fritz_2017}). \citep{Bellhouse_2019} explains in detail the procedure used to derive the SFR for different age bins.

\begin{figure*}\centering
\hbox{\includegraphics[width=0.5\textwidth]{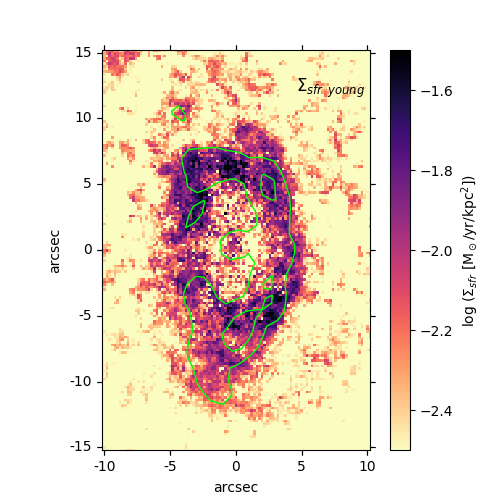}\includegraphics[width=0.50\textwidth]{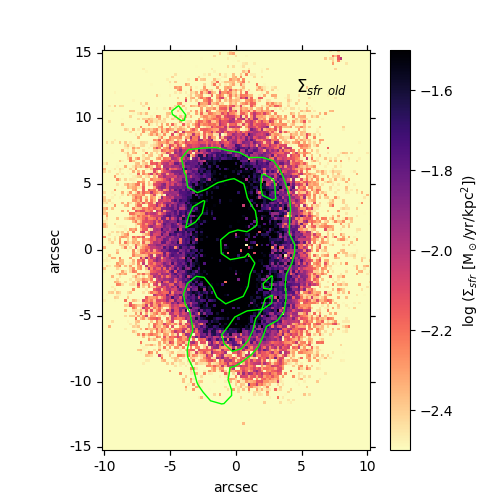}}
\caption{Star formation rate density map of JO201 corresponding to two stellar age ranges, The SFR density for the average stellar ages from .57-to-5.7 $\times$ 10$^8$ yr (young age) is shown in left and from 1.0-to-5.7 $\times$ 10$^9$ yr (old age) is shown in right panel. NUV contours are overlaid (in green) to highlight the cavity with reduced UV flux. We choose same colour scaling for two panels for ease of comparison.}
\label{figure:JO201sfrmap}
\end{figure*}

Fig  \ref{figure:JO201sfrmap} presents the average SFR density maps created for stellar ages from 0.57-to-5.7 $\times$ 10$^8$ yr (we call this as young age)  and 1.0-to-5.7 $\times$ 10$^9$ yr (we call this as old age). The SFR density map created for the younger stellar age is showing a cavity similar to the one seen in UV and supports the hypothesis that there is a reduction in star formation in the central region of JO201 disk compared to the outskirts. The SFR density map created for older ages on the contrary is showing a disk like feature with no cavity. The star formation rate density maps presented here demonstrate that the reduction in star formation happened in the last (5$-$6) $\times$ $10^8$yr. 

We note that our non-parametric approach to reconstruct the star formation as a function of the cosmic time allows us, for spectra of the characteristics and quality such as those we are using for this galaxy, to reach a coarse age resolution which we represent with four age bins.
These were accurately chosen by means of simulations \citep{Fritz_2007}, in such a way that the difference of the spectral features are maximized between different, consecutive, age bins. The reason for this choice is
the impossibility of clearly distinguishing stellar populations of different ages. Hence, the SFR map calculated within each age bin is nothing more but an average value of the stellar mass which was produced during that particular age bin, but whose precise age distribution within this same bin, we are not able to characterize in better detail.
In Fig  \ref{figure:JO201sfrmap} we show the two age bins that are most relevant for our discussion. We stress that we are unable to identify the precise time at which the star formation quenching occurred during the bin (0.57$-$5.7) $\times$ $10^8$yr, we can only assess that it happened at some point during this interval.

\section{Discussion: Star formation suppression due to AGN or stellar bar?} \label{sec:Discussion}

The observed reduction in star formation around the central region of JO201 can in principle be due to the AGN or the stellar bar. Both the feedback from an AGN and the presence of a stellar bar are known to quench star formation in galaxies and it can be difficult to disentangle the relative contribution of each process in suppressing the star formation. We will now discuss in the following the observational pieces of evidence, that can allow us to favor one hypothesis over the other.\\

AGN feedback can inhibit star formation and thereby regulate galaxy evolution as demonstrated in observations and simulations \citep{Sanders_1988,Springel_2005,Dimatteo_2005,Schawinski_2007,Somerville_2008,Hopkins_2008,Schawinski_2010,Wang_2010,Liu_2015,Cheung_2016,Bing_2019}. The outflows and energetic feedback from the AGN can remove the gas from the disk of the galaxy or alternatively induce turbulence working against gas collapse. (also see \citet{Gabor_2014}, who based on simulations have shown that AGN feedback has only a weak effect on gas dynamics of high-redshift disc galaxies.) The jet launched from an accreting black hole influencing galaxy-scale star formation is demonstrated in  recent simulations \citep{Ishibashi_2012,Gaibler_2012,Ishibashi_2014}. The blast wave from the jet can produce an orthogonal bow shock which can push the gas outwards creating a cavity at the centre. The galaxy disk of JO201 is seen almost face on (with a moderate inclination of 54 deg), the AGN is of Seyfert 2 type and the outflow/jet can be tilted from the line of sight of observation. The small connection between the central source and the disk of the galaxy seen in the NUV image can be then due to the effect of geometry in projection. We note that the AGN+composite region in the galaxy disk (see Fig \ref{figure:JO201}) is showing a slight elongation in the east and north west directions which also incidentally coincide with the regions of reduced NUV flux in the horse-shoe area. This can be due to the reduced star formation due to an increasing gas ionization in the direction of an outflow/jet launched from the center of the galaxy disk. If we assume the energy from the AGN is dissipated into the surrounding medium at the speed of light, the time taken by the AGN to create the observed ionization region of size $\sim$ 8.6 kpc in the JO201 disk is $\sim$ 14000 yrs. This should be considered as the minimum time taken by the current phase of AGN activity to ionize the gas. We also note that the AGN in a galaxy typically go through multiple phases of activity that can typically last $\sim$ $10{^5}$ yr \citep{Schawinski_2015}. Therefore it is possible that the star formation suppression is not the effect of a single AGN episode, but due to the net effect of multiple phases of the past AGN activity. There are observational evidences for fossil outflows due to a past strong AGN activity, but now faded in local Universe galaxies \citep{Fluetsch_2019}.


The galaxy scale photoionized narrow line region generated due to the AGN can then extend to several kpc. Therefore, it is possible for the AGN feedback to create the size of ionizing region observed in the disk of JO201. We also point to Fig \ref{figure:JO201bubble}, where the  [OIII] emission line asymmetry map is showing indications of interaction of a possible outflow from AGN with the boundaries of the UV cavity. The scenario of AGN feedback is much more strongly demonstrated in the ALMA CO map in Fig \ref{figure:JO201alma}, where the surrounding areas of the central AGN is devoid of CO (molecular hydrogen). This is also the region of high ionisation temperature as traced by [FeVII] emission line (see Fig \ref{figure:JO201}).\\


The stellar bar in a galaxy can re-distribute the gas making the region close to the bar devoid of fuel for star formation \citep{Combes_1985}. The natural expectation of such a scenario is that the region covered by the length of the stellar bar should be devoid of gas (in molecular, neutral and ionized form) as demonstrated based on a multiwavelength analysis of a face-on barred spiral galaxy Messier 95 \citep{George_2019}. First, we note that, as shown in Fig \ref{figure:JO201bar}, the length of the bar exceeds the size of the cavity. Moreover, the cavity is not devoid of gas, as it is hosting ionized gas as evident from the MUSE emission line maps. This implies the presence of cold gas that had not been redistributed due to a stellar bar prior to being ionized. Hence, the bar could not have suppressed star formation by totally sweeping the cavity of gas. We also note here that the Chandra archive image of JO201 shows no cavity at X-ray wavelengths which supports the scenario of AGN radiative feedback \citep{Ichinohe_2015}.

We conclude that the star formation suppression is the result of recent AGN activity in the central region of JO201 disk over a timescale of $< 5 \times 10^8$yr as revealed from the star formation history map of JO201 shown in Fig \ref{figure:JO201sfrmap}. In contrast, the stellar bar is much older, as testified by its very red color and its length. Unfortunately, SINOPSIS does not provide an exact dating of the bar formation, as any spectrophotometric code loses time resolution for old stellar populations. It should be also noted that JO201 is freshly acquired into the Abell 85 galaxy cluster and started undergoing ram-pressure stripping during the last $\sim$ 1 Gyr \citep{Bellhouse_2019}. The AGN in the center of JO201 may have been triggered as a result of ram-pressure stripping \citep{Poggianti_2017a}.  The most likely explanation for the set of multiwavelength (UV, Optical and CO) observations we have presented is that we are witnessing AGN feedback induced star formation suppression in the central region of JO201. It is interesting to stress that JO201 presents a unique case in which multiple star formation quenching processes appear to be at play. Jellyfish galaxies are undergoing strong ram pressure stripping which is an efficient way of removing gas from the galaxy. The ram-pressure stripping removes the gas from the disk and hence is responsible for the suppression of star formation starting from the outer disk of the galaxy, quenching it "outside-in". Another process (most probably AGN feedback for the reasons outlined above) is operating from the central region of the galaxy and hence quenches the star formation from the inside. JO201 therefore provides a case of "inside-out" and "outside-in" star formation quenching operating in a spiral galaxy. The environment-driven ram-pressure stripping of gas along with the "internal" AGN feedback are contributing to quench the ongoing star formation in the galaxy and will make it at some point join the passive population of red and dead galaxies populating the core of dense clusters.

\section{Summary}

We present a detailed study on the star formation progression on the disk of jelly fish galaxy JO201. Based on a combined analysis of the ultraviolet imaging (UVIT), optical spectroscopy data (MUSE) and CO data (ALMA) we make the following inferences.

\begin{itemize}

\item The galaxy disk of JO201 is characterised by a $\sim$ 8.6 kpc cavity with reduced ultraviolet flux around the AGN. 

\item The Balmer decrement-based A$_{v}$ map of the galaxy disk confirms that the cavity is not due to the effects of localized extinction due to dust but instead is due to the suppression of ongoing/recent star formation.

\item The CO (J$_{2-1}$) map clearly shows a region with no emission in the central region of the galaxy and is situated inside the cavity seen in the UV. This can be considered as an evidence for AGN feedback ionizing the molecular hydrogen and outflows sweeping the gas in its vicinity and thereby inhibiting star formation.

\item The BPT line diagnostics reveals an AGN emission region that matches with the cavity seen in the ultraviolet. The high incidence of AGN at the center of jellyfish galaxies has been suggested to be due to the effect of ram-pressure stripping \citep{Poggianti_2017a}. At the same time, the ram-pressure force enhances the star formation in the outer western side of the disk of JO201. Hence, in this galaxy there is a strong ongoing tussle between the AGN feedback quenching the star formation in the central region and the ram-pressure force (apart from stripping) which compresses the gas in the galaxy disk enhancing star formation.

\item The [FeVII] emission in the central $\sim$ 3 kpc can be explained with the presence of hot gas ($T>10^5$ K) heated either by the AGN continuum or AGN-induced shocks. The MUSE data reveals both a few kpc ongoing AGN outflow and regions of large [OIII] line asymmetry that trace the boundaries of the UV cavity and suggest the presence of a larger spherical outflow or bubble having propagated from the AGN.

\item  The redder (9050-9250 {\AA}) optical image of the galaxy shows the presence of a stellar bar. The stellar bar is prominent at long wavelengths, is long (13 kpc of length) and old, and can be considered as the remnant of (probably secular) evolution of the galaxy before being acquired into the cluster. The  kinematic analysis performed on the red part of the spectrum, including Ca II triplet absorption spectral lines (which trace the  the evolved stellar population) is unable to detect the bar.

\item The star formation history map of JO201 disk demonstrates the existence of a star formation cavity in the last $\sim$ 10$^8$ yr which is absent at older ages. This implies that the cavity seen in UV imaging data is a recent phenomenon.
\end{itemize}

We conclude that the suppression of star formation observed in the central 8.6 kpc of JO201 is due to the effects of AGN feedback happening after infall of the galaxy into the cluster. The observations reported here present a unique example of the combined role of AGN feedback and ram-pressure stripping in the quenching of star formation in spiral galaxies.

\section*{Acknowledgements}

We thank Anna Wolter and Myriam Gitti for discussions on X-ray imaging data of JO201. This publication uses the data from the AstroSat mission of the Indian Space Research  Organisation  (ISRO),  archived  at  the  Indian  Space  Science  Data Centre (ISSDC). Based on observations collected by the European Organisation for Astronomical Research in the Southern Hemisphere under ESO program 196.B-0578 (MUSE). This paper makes use of the following ALMA data: ADS/JAO.ALMA$\#$2017.1.00496.S. ALMA is a partnership of ESO (representing its member states), NSF (USA) and NINS (Japan), together with NRC (Canada), MOST and ASIAA (Taiwan), and KASI (Republic of Korea), in cooperation with the Republic of Chile. The Joint ALMA Observatory is operated by ESO, AUI/NRAO and NAOJ. We acknowledge financial support from PRIN-SKA 2017 (PI L. Hunt). Y.J. acknowledges financial support from CONICYT PAI (Concurso Nacional de Insercion en la Academia 2017), No. 79170132 and FONDECYT Iniciaci\'on 2018 No. 11180558. This project has received funding from the European Research Council (ERC) under the European Union's Horizon 2020 research and innovation programme (grant agreement No. 833824).



\end{document}